# Pressure dependence of upper critical fields in FeSe single crystals

Ji-Hoon Kang[1], Soon-Gil Jung[1], Sangyun Lee[1], Eunsung Park[1], Jiunn-Yuan Lin[2], Dmitriy A Chareev[3,4], Alexander N Vasiliev[5,6,7], and Tuson Park[1]

[1]Department of Physics, Sungkyunkwan University, Suwon 440-746, Republic of Korea

[2]Institute of Physics, National Chiao Tung University, Hsinchu 30010, Taiwan

[3]Institute of Experimental Mineralogy, Russian Academy of Sciences, Chernogolovka, Moscow Region 142432, Russia

[4]Institute of Physics and Technology, Ural Federal University, Mira st. 19, 620002 Ekaterinburg, Russia

[5]Low Temperature Physics and Superconductivity Department, Physics Faculty, Moscow State University, 119991 Moscow, Russia

[6]Theoretical Physics and Applied Mathematics Department, Institute of Physics and Technology, Ural Federal University, Ekaterinburg 620002, Russia

[7]National University of Science and Technology "MISiS", Moscow 119049, Russia

E-mal: prosgjung@gmail.com or tp8701@skku.edu

Fax: +82 31 290 7056

**Abstract**

We investigate the pressure dependence of the upper critical fields ($\mu_0 H_{c2}$) for FeSe single crystals with pressure up to 2.57 GPa. The superconducting (SC) properties show a disparate behavior across a critical pressure where the pressure-induced antiferromagnetic phase coexists with superconductivity. The magnetoresistance for $H//ab$ and $H//c$ is very different: for $H//c$, magnetic field induces and enhances a hump in the resistivity close to the $T_c$ for pressures higher than 1.2 GPa, while it is absent for $H//ab$. Since the measured $\mu_0 H_{c2}$ for FeSe samples is smaller than the orbital limited upper critical field ($H^{orb}_{c2}$) estimated by the Werthamer Helfand and Hohenberg (WHH) model, the Maki parameter ($\alpha$) related to Pauli spin-paramagnetic effects is additionally considered to describe the temperature dependence of $\mu_0 H_{c2}(T)$. Interestingly, the $\alpha$ value is hardly affected by pressure for $H//ab$, while it strongly increases with pressure for $H//c$. The pressure evolution of the $\mu_0 H_{c2}(0)$s for the FeSe single crystals is found to be almost similar to that of $T_c(P)$, suggesting that the pressure-induced magnetic order adversely affects the upper critical fields as well as the SC transition temperature.

## 1. Introduction

Following the discovery of the new superconducting material, LaFeAsO$_{1-x}$F$_x$ with a superconducting transition temperature ($T_c$) of 26 K [1], numerous research results involving iron-based superconductors (IBSs) have been reported. IBSs which share a common structural feature with a Fe*Pn* (*Pn* = pnictogen) or Fe*Ch* (*Ch* = chalcogen) layer [2], can be organized or categorized as various types, such as "1111", "122", "111", or "11"-systems [2, 3]. Chemical doping has a huge effect on the physical properties of IBSs, affecting the electrical resistivity, $T_c$, Neel temperature ($T_N$), upper critical field ($\mu_0 H_{c2}$), and more.

For instance, by substituting fluorine (F) for oxygen (O) in the "1111"-system, SmFeAsO$_{1-x}$F$_x$, the resistivity curves show very different behaviours depending on the substitution ratio [4]. Additionally, the effect of pressure on another "1111"-compound, LaO$_{1-x}$F$_x$FeAs, developed a steep increase in the $T_c$ onset to a maximum value of ~43 K at ~4 GPa [5].

The binary FeSe superconductor which has a $T_c \sim 8$ K [6], is an "11"-system with the simplest crystal structure among the IBSs. In addition to possessing a simple structure, the FeSe superconductor is known to be significantly sensitive to the effects of pressure or doping [7-16]. The upper critical field $\mu_0 H_{c2}$ for IBSs is of great interest, because it provides important information on various superconducting properties, such as coherence length ($\xi$), the pair-breaking mechanism, and anisotropy ($\gamma$) [17-20]. A few studies have been reported on the $\mu_0 H_{c2}$ for FeSe [10, 11, 21-24], stating that along with the orbital effect, a field induced pair-breaking effect or multiband effect also occurs in FeSe. The orbital upper critical field was calculated via the Werthamer-Helfand-Hohenberg (WHH) model that was often used for dirty single band superconductors [25]. Recently, the possibility of Pauli limiting behaviour in FeSe, especially for the *H//ab*, had been reported by high magnetic field measurements [22-24]. In addition, the strong pressure dependence of $\mu_0H_{c2}$ for polycrystalline FeSe was presented by Mizuguchi *et al* [10] and Tiwari *et al* [11] in which they showed a large enhancement of $\mu_0H_{c2}(0)$ from 28 T (0 GPa) to 50 T (1.48 GPa) and from 26.7 T (0 GPa) to 47.5 T (1.98 GPa), respectively.

Even though FeSe has the simplest crystal structure among IBSs and is significantly affected by pressure, detailed studies on the $\mu_0H_{c2}$ of FeSe single crystals under pressure have not yet been reported due to difficulty in the synthesis of high-quality *c*-axis-oriented FeSe single crystals [10, 24, 26-28]. In this paper, we present the pressure dependence of upper critical fields ($\mu_0H_{c2}$) for high-quality *c*-axis-oriented FeSe single crystals via temperature dependence of resistivity measurements under pressure and magnetic fields which are applied parallel (*H//ab*) and perpendicular (*H//c*) to the *ab*-plane.

## 2. Experiments

High-quality $FeSe_{1-x}$ (*x* = 0.04 ± 0.02) single crystals with *c*-axis orientation were synthesized via $AlCl_3$/KCl flux, where the synthetic details were described elsewhere [26, 29]. A few FeSe layers of the single crystals were easily exfoliated using adhesive tapes and were used in this study. Four FeSe samples (FeSe#1 and #2, and FeSe#3 and #4) from the same batch were used such that the applied magnetic field was parallel (*H//ab*) and perpendicular (*H//c*) to the *ab*-plane of the crystals. Quasi-hydrostatic pressure was applied via a clamp-type piston-cylinder pressure cell [30] with Daphne oil 7373 as the pressure-transmitting medium. Transport measurements were performed by a physical property measurement system (PPMS 9 T, Quantum Design) in which the pressure cell was cooled to 1.8 K and magnetic fields up to 9 T were applied to the samples. Electrical resistivity measurements were carried out using a standard four-probe method in which the electrical contact was made by using silver epoxy. In order to calculate the precise pressure applied to the samples, a change in the SC transition temperature of lead (Pb), which was contained together within the pressure cell, was resistively measured.

## 3. Results and discussion

Figure 1 presents the in-plane resistivity ($\rho_{ab}$) as a function of temperature at ambient pressure. In the normal state, the resistivity curve exhibits a metallic behavior, where $\rho_{ab}$ decreases with decreasing temperature and the residual resistivity ratio ($RRR = \rho_{ab(300K)}/\rho_{ab(13K)}$) is ~11.25±0.75 for four FeSe samples that are presented in this work. A small resistivity anomaly near 80 K is shown, which is due to a tetragonal to orthorhombic structural transition ($T_s$) [31-33]. The top left inset shows a scanning electron microscopy (SEM) image of the fabricated FeSe single crystal, showing a plate-like shape. The bottom right inset shows the criterion of the superconducting transition temperature ($T_c$) used in this study, which was determined from the peak in the temperature derivative of the resistivity, $d\rho_{ab}/dT$.

The temperature dependence of $\rho_{ab}$ of FeSe#2 for pressures up to 2.57 GPa is presented in figure 2(a), representatively. The top left inset of figure 2(a) is an enlarged view of the $\rho_{ab}(T)$ near $T_c$, which is shown to highlight the low temperature behaviour with external pressures.

Structural ($T_s$) and magnetic transition temperatures ($T_N$) were determined by the point of the local minimum in d$\rho_{ab}$/d$T$ curve, as shown in the bottom right inset of figure 2(a). When subjected to pressure, the superconducting transition width ($\Delta T_c$) becomes sharper up to 1.0 GPa and broadened at higher pressure, which is consistent with prior studies [10, 11, 14]. The sharper transition indicates that the pressure affected the samples in a way which strengthened their superconductivity. However, at 1.0 GPa a slightly anomalous behaviour was observed at the start of the superconducting transition due to the appearance of magnetic ordering. When antiferromagnetic (AFM) ordering is introduced, the superconducting transition becomes broader, as shown for pressures higher than 1.0 GPa [12-14], due to competition between superconductivity and antiferromagnetism.

The pressure evolution of phase diagram for the FeSe single crystals is described in figure 2(b), where superconducting ($T_c$), structural ($T_s$), and magnetic ($T_N$) transition temperatures are plotted based on the resistivity measurements [34]. The $T_c(P)$ showed local minima at the critical pressure ($P_c$) between 1.0 and 1.3 GPa because of the emergence of the magnetic state around these pressures [12-14], which is a unique feature in the FeSe superconductor compared with other IBSs. For comparison, we also plotted the magnetic phase transition temperature obtained from $\mu$SR measurements [12], which are in good agreement with our results.

Figure 3(a) shows the temperature dependences of normalized $\rho_{ab}$ for FeSe#1 and FeSe#2 at selected pressures of 0.78, 1.29, 1.70, and 2.35 GPa. In the FeSe#1, interestingly, the hump is slightly developed by applied pressure of 1.29 GPa at which the magnetic state is emerged [12-14], and the largest hump is presented at 1.70 GPa. The anomalous hump appears below the $T_c$ onset of FeSe#2, which is clearly observed in $\rho_{ab}(T)$ curves at 1.70 GPa where the dashed line is guide for the eyes. In addition, the hump in the FeSe#2 is also induced above 1.29 GPa by applying magnetic fields, although any signature for the hump is not observed at zero field. The $\rho_{ab}(T)$'s for the FeSe#1 (*H//ab*) and #2 (*H//c*) at 1.70 GPa are representatively shown in figures 3 (b) and (c), respectively, because the hump is the most prominent at that pressure.

Recently, pressure-induced resistive hump around $T_c$ in FeSe had beed reported because of pressure-induced antiferromagnetism [14]. In this study, the hump is also developed by application of pressure, which showed the sample dependence and the dependence on the direction of applied magnetic fields, as shown in figures 3 and 4. The sample dependence is possibly related with an inhomogeneity of crystals, such as due to selenium (Se) deficiencies. However, the field-direction dependence of the hump cannot be explained by an inhomogeneity of samples. The hump is sensitive for *H//c* above the $P_c$, as shown in figure 3(c) and figures 4(g) and (h). On the other hand, the hump is hardly noticeable for *H//ab*, as shown in figures 4(c) and (d). These results suggest that the hump is closely related to the pressure-induced magnetic state and magnetic field induces an AFM order parameter by suppressing superconductivity because the AFM order in the FeSe is associated with the nearest-neighbor Fe ions in the *ab*-plane [12].

The $\rho_{ab}(T)$ curves for different pressures in *H//ab* (FeSe#3) and *H//c* (FeSe#4) are shown in figures 4(a), (b), (c), (d) and (e), (f), (g), (h), respectively. For *H//ab* and *H//c*, the magnetoresistance (MR) for both crystals is very different: for *H//ab*, the normal state resistivity is almost independent of the applied magnetic field, while it is strongly dependent on the field for *H//c*. The anisotropic positive MR in FeSe was also reported in other recent works, which was ascribed to a possible Fermi surface reconstruction [23, 35-38].

The normalized temperature ($t = T/T_c$) dependence of the upper critical field ($\mu_0 H_{c2}$), for both directions $H//ab$ (FeSe#3) and $H//c$ (FeSe#4) are plotted in figures 5(a), (b), (c) and (d), (e), (f), respectively, where temperature is normalized by the SC $T_c$ in order for comparison among different pressures with different $T_c$. The $T_c$ at each magnetic field was determined from the peak in the $d\rho_{ab}/dT$ curves, as illustrated in figure 1. The $\mu_0H_{c2} - T$ curves for IBSs usually exhibit puzzling curvatures and behaviours which are ascribed to spin-paramagnetic and/or multiband effects [17, 18, 20, 39]. First, we calculated the orbital upper critical field via the Werthamer-Helfand-Hohenberg (WHH) method using the relationship, $\mu_0H^{orb}_{c2}(0) = -0.693 \times T_c \times (dH_{c2}/dT)_{T=T_c}$ for the dirty limit [25]. The $\mu_0H^{orb}_{c2}(t)$s for 0.41, 1.72, and 2.43 GPa are representatively plotted as the dashed lines in figure 5, which show larger value than the measured $\mu_0H_{c2}(t)$. Accordingly, we considered the contribution of Pauli spin-paramagnetic (PSP) effect by the use of the Maki parameter, indicated by the solid lines at each pressure in figure 5, which show much suppression at low temperatures compared to the dashed lines. The temperature dependence of the upper critical fields is given by

$$H_{c2}(t) = \frac{H^{orb}_{c2}(0)}{\sqrt{1+\alpha^2}}(1-t^2), \qquad (1)$$

where α is the Maki parameter [40], indicating that not only the orbital effect but also the PSP effect has an influence on the pair-breaking mechanism through the entire pressure region. In addition, the deviation between $\mu_0H^{orb}_{c2}(t)$ and $\mu_0H_{c2}(t)$ becomes large with an increase in pressure for $H//c$, indicating that the PSP effect is more important for pair-breaking along this direction. The information for the $T_c$, upper critical fields and Maki parameter at each pressure for the FeSe samples is summarized in table 1. The upper critical fields at 0 K and ambient pressure are 25.5 T (FeSe#1) and 22.5 T (FeSe#3) for $H//ab$ and 13.3 T (FeSe#2) and 10.5 T (FeSe#4) for $H//c$, which are comparable to the upper critical fields obtained from high field measurements in references [22, 23]. In addition, a large enhancement of $H_{c2}(0)$ under pressure is similar to that of $H_{c2}(0)$ under pressure for polycrystalline FeSe [10, 11].

Figures 6(a) and (b) present $\mu_0 H_{c2}$ at $t = 0$ and 0.9 as a function of pressure for FeSe single crystals along both directions $H//ab$ and $H//c$, respectively. The points for $t = 0$ were estimated from equation (1) mentioned above, whereas the points for $t = 0.9$ were measured from the $\rho_{ab}(T)$ in magnetic fields. The pressure dependence of $\mu_0 H_{c2}(P)$ is similar to $T_c(P)$, showing a local minimum at 1.3 GPa where a magnetic state is induced. For $t = 0$ and 0.9, a small drop in both $\mu_0 H_{c2}(t)//ab$ and $\mu_0 H_{c2}(t)//c$ appeared at the highest pressure (=2.57 GPa) in this study, even though $T_c$ is enhanced. The tetragonal to orthorhombic structural transition is extrapolated to be zero Kelvin at this pressure [41], indicating an interplay between the two order parameters [42].

The anisotropy ($\gamma_H$) in the upper critical fields is estimated from the FeSe#3 and #4 for the same pressures, because their physical properties, such as residual resistivity near $T_c$, $T_c(P)$, and no hump in the $\rho_{ab}(T)$ at zero-field, are very similar to each other. The $\gamma_H$ at both $t = 0$ and 0.9 decreases with increasing pressure, as shown in figure 6(c). The pressure dependence of the Maki parameter ($\alpha$) obtained from the equation (1) is presented in figure 6(d). Contrary to $H//ab$ where $\alpha$ is almost independent of pressure, $\alpha$ for $H//c$ increases with increasing pressure, indicating that the PSP effect for $H//c$ becomes increasingly important as pressure is increased. Even though our results are estimated from the low-field data, the estimated $\alpha$ at ambient pressure is in good agreement with the high-field results for high-quality FeSe single crystals [23, 43]. The strong PSP effect was observed in many IBSs in which $\alpha$ has a larger value for $H//ab$ than $H//c$ [17, 18, 44], and the FeSe also showed the similar trend with other IBSs. For $H//c$, the negligible PSP effect had been reported in FeSe single crystals [22-24], while a large

PSP effect with $\alpha > 1$ was presented by a high-field measurement for $H//ab$ [22]. Further study at higher fields in pressure is needed to make a more definitive conclusion on the anisotropic Maki parameter.

## 4. Conclusions

In conclusion, we have studied the effects of pressure on the upper critical field of FeSe single crystals by measuring the in-plane electrical resistivity in pressures up to 2.57 GPa and magnetic fields up to 9 T applied parallel ($H//ab$) and perpendicular ($H//c$) to the $ab$-plane. At low pressures, where only superconductivity exists, $T_c$ was enhanced and the superconducting transition width became sharper with increasing pressure. However, near 1.0 GPa where the magnetic phase was induced, anomalous resistivity and superconducting critical temperature behaviours were observed. From the resistivity curves of the FeSe samples, we were able to measure the superconducting, structural, and magnetic phase transition temperatures to build a comprehensive pressure phase diagram, which is comparable with other reports [14, 41]. Similar pressure dependence between $\mu_0 H_{c2}(P)$ and $T_c(P)$ underlines that the pressure-induced magnetic state not only influences the critical temperature but also the upper critical field. The anisotropy ($\gamma_H$) in the upper critical fields decreases with increasing pressure. Furthermore, Pauli spin-paramagnetic effect for $H//c$ shows a linear-in-pressure dependence, whereas that for $H//ab$ is almost independent of pressure.

**Note Added**

Since this work was submitted, the similar pressure dependence of $H_{c2}$ of FeSe from the measurements of the inter-plane electrical resistivity was reported in reference [45].

**Figure Captions**

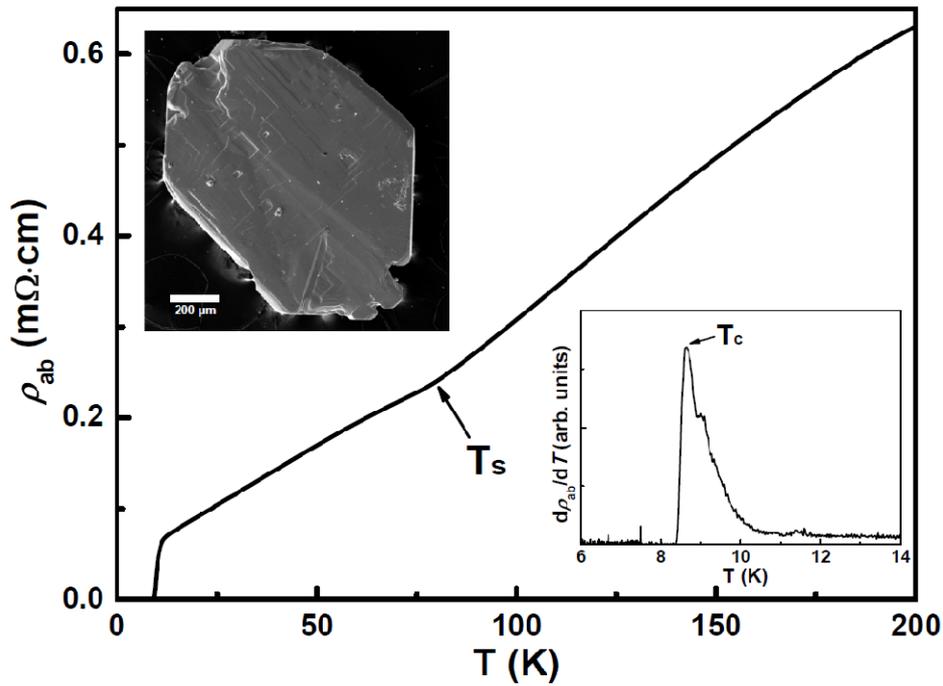

Figure 1. Temperature dependence of the in-plane resistivity ($\rho_{ab}$) for FeSe single crystals at ambient pressure. The structure transition ($T_s$) from tetragonal to orthorhombic structure occurs around 80 K indicated by the arrow. The top left inset shows the SEM image of the FeSe single crystal. The bottom right inset shows the first temperature derivative of $\rho_{ab}$ as a function of temperature, where the superconducting transition temperature ($T_c$) was assigned as the peak in $d\rho_{ab}/dT$.

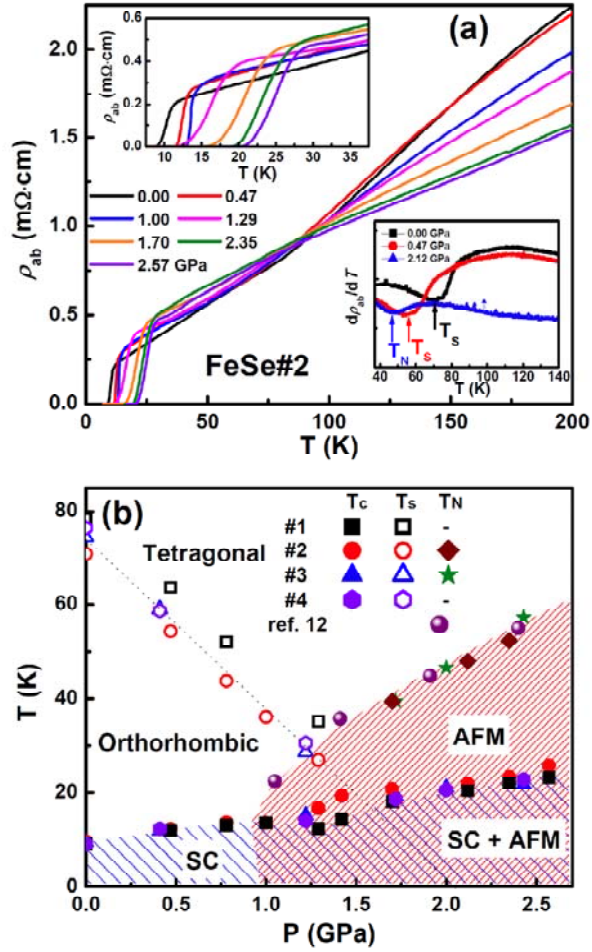

Figure 2. (a) Temperature dependences of in-plane resistivity ($\rho_{ab}$) for the FeSe#2 in pressures up to 2.57GPa. The top left inset is an enlarged view of the $\rho_{ab}(T)$ near $T_c$, which shows a broadening of the superconducting transition for $P \geq 1.29$ GPa. The bottom right inset displays the first temperature derivative of $\rho_{ab}$, where the structural transition temperature ($T_s$) and magnetic transition temperature ($T_N$) were assigned as the minimum in $d\rho_{ab}/dT$. (b) The pressure phase diagram of the FeSe single crystals. The superconducting ($T_c$), structural ($T_s$), and magnetic ($T_N$) transition temperatures from the resistivity measurements are plotted. The dotted line is a guide for the eyes and SC and AFM denote superconducting and antiferromagnetic states.

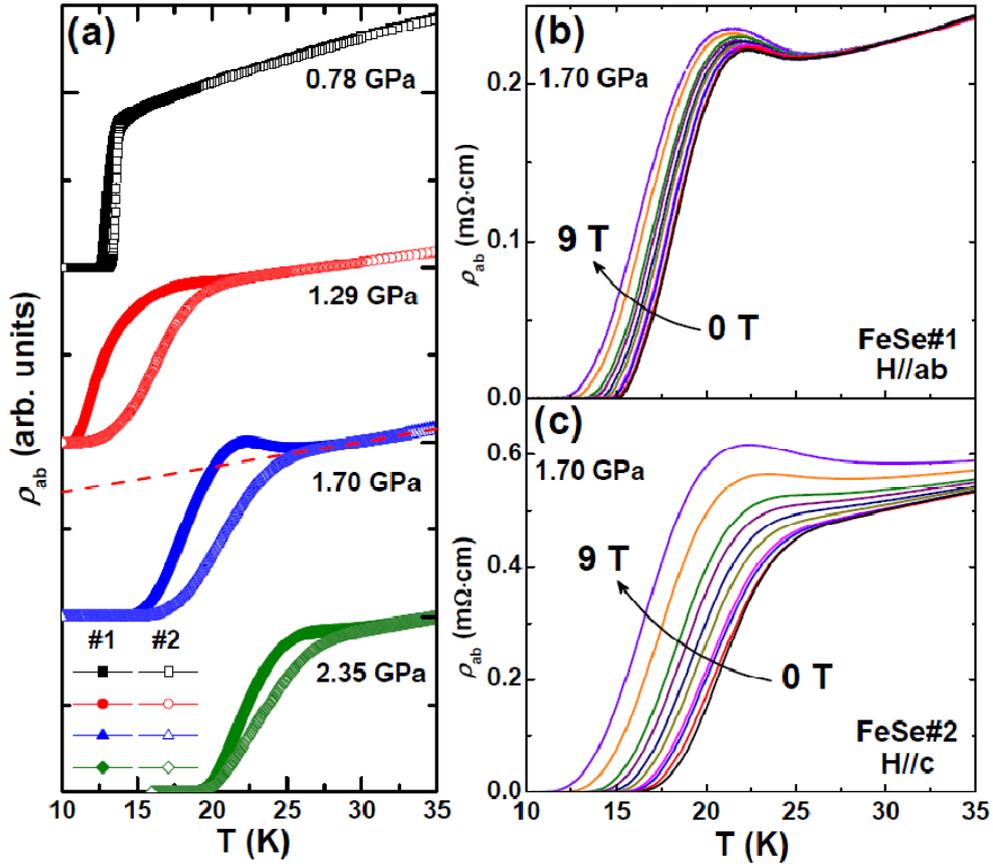

Figure 3. (a) Temperature dependence of the in-plane resistivity ($\rho_{ab}$) of FeSe single crystals at $P$ = 0.78, 1.29, 1.70, and 2.35 GPa, where $\rho_{ab}(T)$'s at 0.78, 1.29, 1.70, and 2.35 GPa are normalized by the resistivity at 18, 27, 30, and 35 K, respectively, for comparison. A hump appears at $P \geq 1.29$ GPa for FeSe#1. The dashed line at 1.70 GPa is a guide for the eyes. The resistivity $\rho_{ab}(T)$ of FeSe single crystals at 1.70 GPa is displayed for magnetic fields applied parallel ($H//ab$) and perpendicular ($H//c$) to the $ab$-plane in (b) and (c), respectively, where the humps become more noticeable with increasing magnetic field.

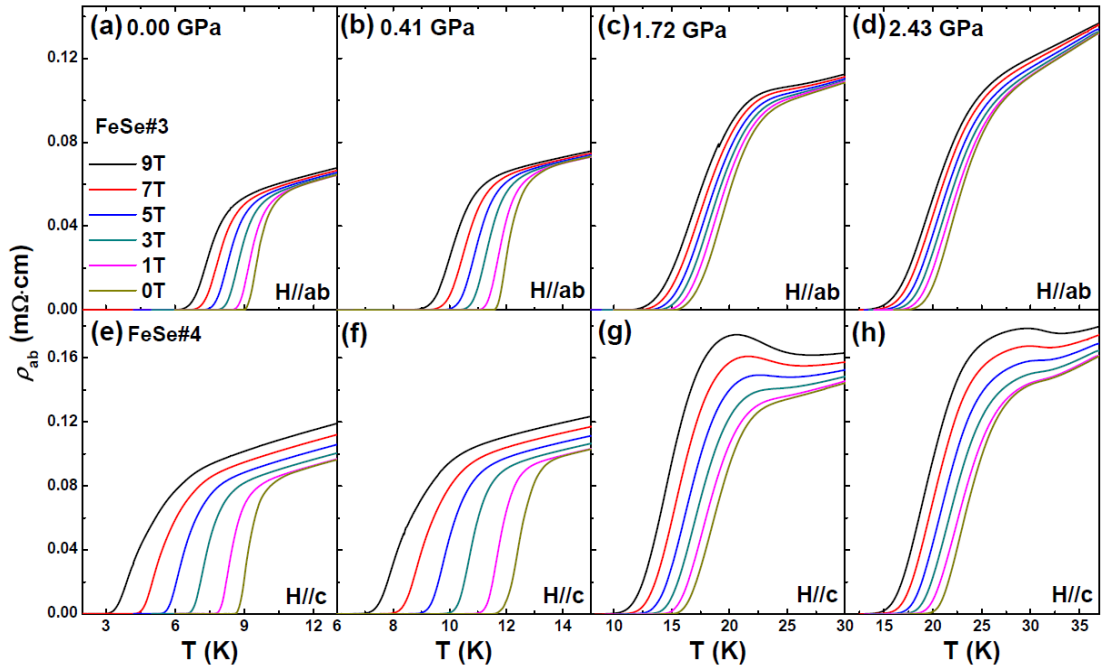

Figure 4. $\rho_{ab} - T$ curves for FeSe single crystals in magnetic fields are selectively shown at 0, 0.41, 1.72, and 2.43 GPa. The magnetic fields are applied parallel ($H//ab$) [(a), (b), (c), (d)] to the $ab$-plane for FeSe#3 single crystal, while the fields are applied perpendicular ($H//c$) [(e), (f), (g), (h)] to the $ab$-plane for FeSe#4 single crystal, Both FeSe#3 and #4 single crystals were obtained from the same batch and show similar superconducting as well as normal state properties at ambient pressure and zero field.

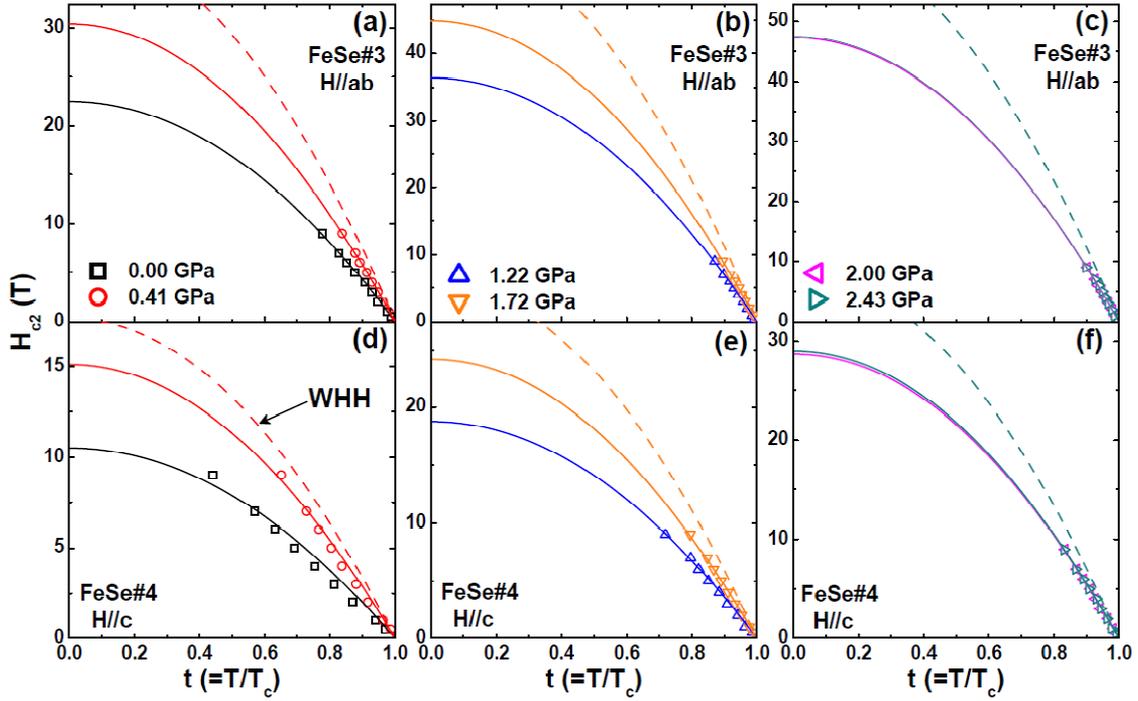

Figure 5. Upper critical field ($\mu_0 H_{c2}$) of FeSe single crystals as a function of reduced temperature ($t = T/T_c$) at different pressures, where magnetic fields are applied parallel ($H//ab$) [(a), (b), (c)] and perpendicular ($H//c$) [(d), (e), (f)] to the *ab*-plane. The dotted lines describe the upper critical fields at 0.41GPa [(a) and (d)], 1.72GPa [(b) and (e)], and 2.43GPa [(c) and (f)] estimated by the WHH model, and the solid lines show the upper critical fields plotted from the WHH model with including the Pauli-limiting effect.

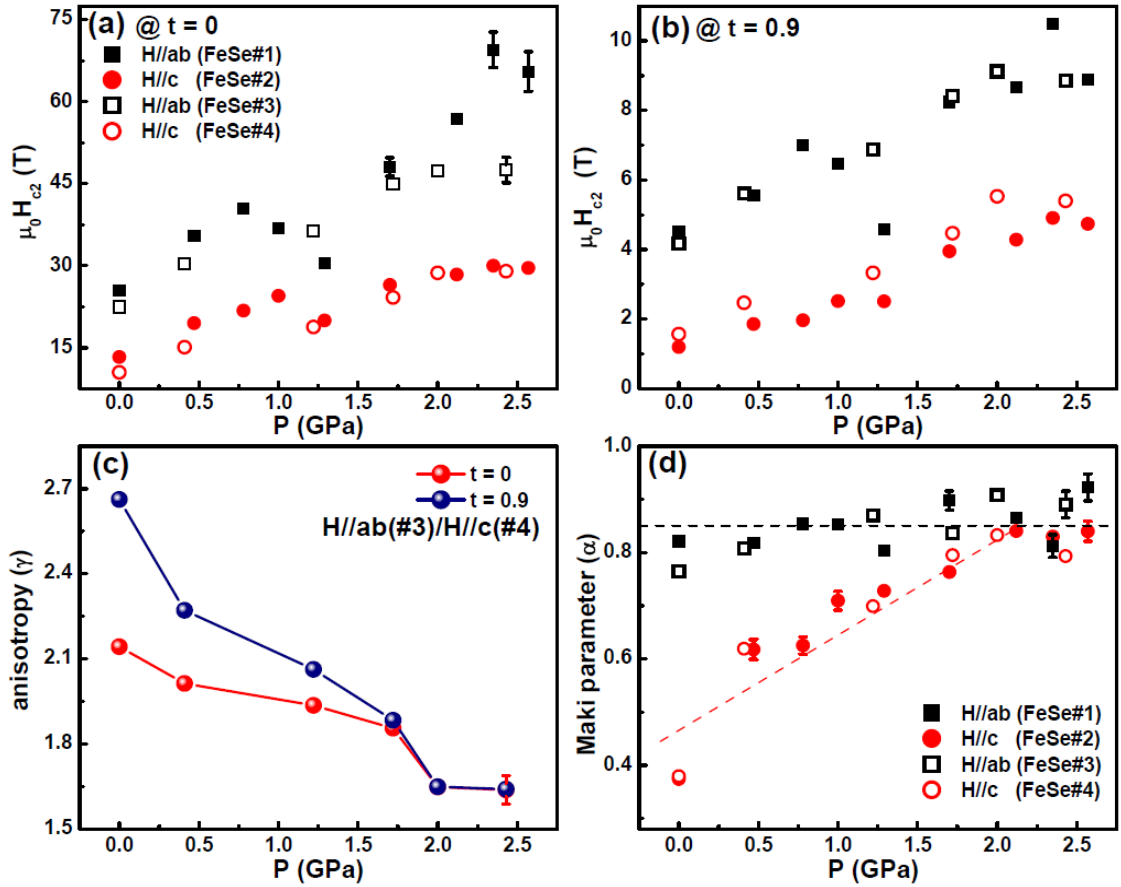

Figure 6. Upper critical field ($\mu_0 H_{c2}$) as a function of pressure for FeSe single crystals for both $H//ab$ and $H//c$ field directions at (a) $t = 0$ and (b) 0.9. (c) Anisotropy ($\gamma_H$) of the upper critical fields, $\mu_0 H^{ab}_{c2}/\mu_0 H^{c}_{c2}$, is plotted at $t = 0$ and 0.9, which decreases with increasing pressure. (d) Pressure dependences of Maki parameter ($\alpha$). The value of $\alpha$ that reflects the influence of the orbital and spin-paramagnetic effect on the pair-breaking mechanism is almost constant over the whole pressure regions for $H//ab$, whereas it increases rapidly with pressure for $H//c$.

Table 1. Summary of the upper critical field for the FeSe single crystals. ($P$: pressure, $T_c$: superconducting critical temperature, $(dH_{c2}/dT)_{T=T_c}$: slope of $H_{c2}$ near $T_c$, $\mu_0 H^{orb}_{c2}$: orbital upper critical field $\mu_0 H_{c2}$: upper critical field, $\alpha$: Maki parameter).

| $P$ (GPa) | $T_c$ (K) | | $(dH_{c2}/dT)_{T=T_c}$ (T/K) | | $\mu_0 H^{orb}_{c2}(0)$ (T) | | $\mu_0 H_{c2}(0)$ (T) | | $\alpha$ | |
|---|---|---|---|---|---|---|---|---|---|---|
| | *H//ab* (#1) | *H//c* (#2) | *H//ab* (#1) | *H//c* (#2) | *H//ab* (#1) | *H//c* (#2) | *H//ab* (#1) | *H//c* (#2) | *H//ab* (#1) | *H//c* (#2) |
| 0.00 | 8.64 | 9.66 | -5.52 | -2.12 | 33.1 | 14.2 | 25.5 | 13.3 | 0.82 | 0.37 |
| 0.47 | 11.42 | 12.31 | -5.86 | -2.69 | 45.9 | 22.9 | 35.5 | 19.5 | 0.82 | 0.62 |
| 0.78 | 12.99 | 13.66 | -5.97 | -2.72 | 53.3 | 25.7 | 40.5 | 21.8 | 0.85 | 0.63 |
| 1.00 | 13.55 | 13.6 | -5.19 | -3.19 | 48.4 | 30.0 | 36.8 | 24.5 | 0.85 | 0.71 |
| 1.29 | 12.38 | 16.73 | -4.66 | -2.13 | 39.1 | 24.7 | 30.5 | 20.2 | 0.80 | 0.71 |
| 1.70 | 18.16 | 20.73 | -5.20 | -2.32 | 64.6 | 33.3 | 48.1 | 26.5 | 0.90 | 0.76 |
| 2.12 | 20.44 | 21.89 | -5.35 | -2.45 | 75.1 | 37.1 | 56.8 | 28.4 | 0.87 | 0.84 |
| 2.35 | 21.95 | 23.37 | -6.00 | -2.41 | 89.5 | 39.0 | 69.5 | 30.0 | 0.81 | 0.83 |
| 2.57 | 23.27 | 25.75 | -5.68 | -2.17 | 89.1 | 38.7 | 65.5 | 29.6 | 0.92 | 0.84 |
| | *H//ab* (#3) | *H//c* (#4) | *H//ab* (#3) | *H//c* (#4) | *H//ab* (#3) | *H//c* (#4) | *H//ab* (#3) | *H//c* (#4) | *H//ab* (#3) | *H//c* (#4) |
| 0.00 | 9.54 | 9.05 | -4.27 | -1.79 | 28.2 | 11.2 | 22.5 | 10.5 | 0.76 | 0.38 |
| 0.41 | 11.98 | 12.36 | -4.71 | -2.07 | 39.1 | 17.7 | 30.4 | 15.1 | 0.81 | 0.62 |
| 1.22 | 14.95 | 13.98 | -4.57 | -2.37 | 47.3 | 23.0 | 36.4 | 18.8 | 0.81 | 0.70 |
| 1.72 | 19.30 | 18.58 | -4.38 | -2.40 | 58.6 | 30.9 | 44.9 | 24.2 | 0.84 | 0.80 |
| 2.00 | 21.08 | 20.59 | -4.37 | -2.63 | 63.8 | 37.5 | 47.3 | 28.7 | 0.91 | 0.83 |
| 2.43 | 22.11 | 22.88 | -4.22 | -2.33 | 64.7 | 37.0 | 47.5 | 29.0 | 0.89 | 0.79 |